\documentclass[prl, twocolumn, superscriptaddress, aps]{revtex4}
 
\usepackage{amsmath,amsfonts, amssymb, amsthm, dsfont}
\usepackage[mathscr]{euscript}
\usepackage{bm}
\usepackage{mathrsfs}
\usepackage{graphicx}
\usepackage{verbatim}
\usepackage{hyperref}

\renewcommand{\vec}[1]{{\mathbf #1}}

\newcommand{\vx}{{\vec{x}}}
\newcommand{\vy}{{\vec{y}}}

\begin{document}

\title{The emergence of 3+1D Einstein gravity from topological gravity theory}
\author{Zheng-Cheng Gu}
\affiliation{Department of Physics, The Chinese University of Hong Kong, Shatin, New Territories, Hong Kong, China}
\date{{\small \today}}

\begin{abstract}
Quantum field theory successfully explains the origin of all fundamental forces except gravity due to the renormalizability problem. In this paper, we proposed a topological scenario to understand this puzzle. First, we proposed a $3+1$D topological (quantum) gravity theory which is renormalizable, and it can be regarded as a straightforward generalization of Edward Witten's Chern-Simons theory approach to $2+1$D topological gravity. Then, we showed that the (vacuum) Einstein-Cartan equation and classical space-time naturally emerge from topological (quantum) gravity via loop condensation. The second step is a unique feature in $3+1$D and it might even naturally explain why our space-time is four dimensional. %Experimentally measurable low energy predictions are also discussed.    
\end{abstract}

\maketitle

\textit{Introduction} -- Recent discovery of gravitational wave by LIGO\cite{LIGO} verifies Einstein's theory of gravity and brings back the old paradox between general relativity and quantum mechanics. Naively, it has been argued that the absolute time in quantum mechanics is intrinsically inconsistent with diffeomorphism invariance and that it is impossible to construct a gravitational theory that can be consistently quantized. 

On the other hand, the modern perspective of continuum field theory based on the concept of renomalization group(RG) suggests that any meaningful continuum field theory must emerge as an effective theory from an underlying RG fixed point, hence it must be a renormalizable quantum field theory(QFT). Indeed, the well known standard model is controlled by a conformal field theory(CFT) fixed point in the asymptotic freedom limit. Therefore, the essential task of defining a quantum theory of gravity becomes defining new types of RG fixed point that can reproduce Einstein's gravity in the semi-classical limit. Recent development in ADS/CFT correspondence conjecture provides us a novel example of defining $d+1$-dimensional ADS-space quantum gravity in terms of $d$-dimensional CFT\cite{ADSCFT1,ADSCFT2}. Nevertheless, the ADS/CFT correspondence does not work in De Sitter space. Hence a much more general physical concept and mathematical framework of understanding quantum gravity are very desired. The so-called loop quantum gravity(LQG) is a very interesting attempt along this direction\cite{LQG1,LQG2,LQG3,LQG4,LQG5}.

Three decades ago, Edward Witten proposed to use Chern-Simons theory to reformulate $2+1$D Einstein's gravity\cite{CS1} and a consistent quantum gravity theory can be defined (at least perturbatively)in the absence of matter fields(with or without cosmological constant term). Although $2+1$D gravity is somewhat trivial due to the absence of propagating gravitational wave and vanishing of space-time curvature, it still provides us a concrete example of understanding quantum gravity in terms of a topological quantum field theory(TQFT). Moreover, according to the correspondence between Chern-Simons theory and CFT, the ADS3/CFT2 correspondence conjecture can be understood in a very natural way\cite{ADSCFT3}.

Nevertheless, the TQFT approach can not be easily generalized into $3+1$D due to the following difficulties. (a) Einstein's gravity in $3+1$D contains propagating mode -- the gravitational wave, therefore it is obviously not a TQFT in the usual sense. (b) Our knowledge of higher dimensional TQFT is very limited and there is no Chern-Simons like action in $3+1$D. 

Thanks to the recent development of the classification of topological phases of quantum matter in higher dimensions\cite{Chenlong,Wenscience,cobordism,Wencoho}, new types of TQFT have been discovered in $3+1$D to describe the so-called three-loop-braiding statistics. In this paper, we argued that such a TQFT is closely related to Einstein gravity. In particular, we conjectured that gravitational wave will disappear at an extremely high energy scale and $3+1$D quantum gravity is indeed controlled by a TQFT RG fixed point.  At an intermediate energy scale, Einstein gravity and classical space-time can emerge via loop(flux lines) condensation. In the loop condensed phase, the quantum fluctuation is controlled by a small parameter $\theta$ and the theory is still power-counting renormalizable. In the semi-classical limit, we can derive the same equation of motion as Einstein-Cartan equation(in the absence of matter fields). Furthermore, our theory predicts the non-commutative geometry between spin connection $\omega$ and curvature tensor $R$.

\textit{$3+1$D Topological gravity} --
In Edward Witten's pioneer work for $2+1$D quantum gravity, he pointed out that the Einstein-Cartan action in $2+1$D can be regarded as a Chern-Simons action $\int Tr[A\wedge (dA+A\wedge A)]$ where $A$ is the gauge connection of Poincare group $ISO(2,1)$($SO(3,1)$ or $SO(2,2)$ for nonzero cosmological constant case). However, he further argued that since there is no $\int Tr[ A\wedge A\wedge (dA+A\wedge A)]$ type topological quantum field theory(TQFT) in $3+1$D, the corresponding Einstein-Cartan action can not be regarded as a TQFT and in fact it is even not a well defined renormalizable QFT. The tremendous efforts on super gravity theory\cite{SUGRAV1,SUGRAV2} and ADS/CFT correspondence conjecture all aim to developed a well defined QFT description for gravity.(We hesitate to mention super string theory here since its relevant part to physics still relies on super gravity and ADS/CFT correspondence.) 

Instead of using super symmetry(SUSY) and ADS/CFT correspondence to define and understand quantum gravity, here we attempt to consider the problem from a different angle. We would like to ask: Is there any TQFT in $3+1$D that is closely related to the Einstein gravity, e.g., can we realize Einstein gravity through a proper phase transition from a TQFT? There are several advantages in TQFT approach to quantum gravity, as having already been demonstrated in the $2+1$D case. First, it is manifested renormalizable and super symmetry is not necessary. Second, it can handle general cases with or without cosmological constant. 
%This is a unique advantage comparing to the SUSY and ADS/CFT correspondence approach, where an ADS background usually can not be avoided. 
%Finally, the TQFT approach even has the potential to remove UV singularities in black hole and big bang theory.

Surprisingly, recent development in condensed matter physics indicates that $\int Tr[ A\wedge A\wedge (dA+A\wedge A)]+\int Tr(B \wedge F)$ type actions\cite{aada1,aada2,aada3} do exist and might serve as the most general $3+1$D TQFT that describes nontrivial three-loop-braiding statistics\cite{loop1,loop2}. For discrete gauge group, they are known as Dijkgraaf-Witten theories\cite{DWmodel}. Now let us generalize the above action into Poincare group and define the following topological gauge theory: 
\begin{widetext}
\begin{eqnarray}
S_{\rm{Top}}&=&\frac{1}{2}\int \epsilon_{a
bcd} e^a \wedge e^b \wedge R^{cd}+\int   B_{ab} \wedge R^{ab}+\int  \tilde{B}_a \wedge T^{a}\nonumber\\
&=&\frac{1}{4}\int d^4 x\epsilon^{\mu\nu\rho\sigma}\epsilon_{a
bcd}{e_\mu}^a {e_\nu}^b {R_{\rho\sigma}}^{cd}+\frac{1}{4}\int d^4 x\epsilon^{\mu\nu\rho\sigma}B_{\mu\nu ab}{R_{\rho\sigma}}^{ab}+\frac{1}{4}\int d^4 x\epsilon^{\mu\nu\rho\sigma}\tilde{B}_{\mu\nu a}{T_{\rho\sigma}}^{a}
\end{eqnarray}
\end{widetext}
Here $B,\tilde{B}$ are 2-form gauge fields which were first introduced in usual topological $BF$ theory\cite{bf1,bf2,bf3}, and $R,T$ are the usual curvature and torsion tensors:
\begin{eqnarray}
{R_{\rho\sigma}}^{cd}&=& \partial_\rho{\omega_{\sigma}}^{cd}-\partial_\sigma{\omega_{\rho}}^{cd}+{\omega_{\rho}}^{ce}{\omega_{\sigma e }}^{d}
-{\omega_{\sigma}}^{ce}{\omega_{\rho e}}^{d},\nonumber\\
{T_{\mu\nu}}^a&=& \partial_\mu {e_\nu}^a-\partial_\nu {e_\mu}^a+{\omega_\mu}^{ab}e_{\nu b}-{\omega_\nu}^{ab}e_{\mu b}\label{action}
\end{eqnarray}
The first term in the above action is the usual Einstein-Cartan action. It is easy to verify that such a topological action is invariant under local Lorentz symmetry transformation. Interestingly, the total action is actually invariant under the whole local Poincare symmetry transformation, if we properly define the gauge transformation of translational symmetry for 2-form gauge fields:  
\begin{eqnarray}
{e_\mu}^a &\rightarrow & {e_\mu}^a+D_\mu f^a \equiv {e_\mu}^a+\partial_\mu f^a+{\omega_\mu}^{ab} f_b \nonumber\\
\tilde{B}_{\mu\nu a} &\rightarrow & \tilde{B}_{\mu\nu a}-\epsilon_{a
bcd} f^b {R_{\mu\nu}}^{cd} \nonumber \\
B_{\mu\nu ab} &\rightarrow & B_{\mu\nu ab}-\frac{1}{2}(\tilde{B}_{\mu\nu a}f_b-\tilde{B}_{\mu\nu b}f_a)\label{gauge}
\end{eqnarray}
We note that the usual first order Einstein-Cartan action is not invariant under the above gauge transformation, and that's why it is not a well defined TQFT in $3+1$D.
In addition, we can also define the following gauge transformation for 2-form gauge fields $\tilde{B}_{\mu\nu a}$ and $B_{\mu\nu ab}$:
\begin{eqnarray}
{\tilde{B}_{\mu\nu a}} &\rightarrow&   {\tilde{B}_{\mu\nu a}}+\partial_\mu {\tilde{\xi}_{\nu a}}-\partial_\nu {\tilde{\xi}_{\mu a}}+{\omega_{\mu a}}^{b}\tilde{\xi}_{\nu b}-{\omega_{\nu a}}^{b}\tilde{\xi}_{\mu b}\nonumber \\ {B_{\mu\nu ab}} &\rightarrow& {B_{\mu\nu ab}}- \frac{1}{2}[({\tilde{\xi}_{\mu a}} {e_{\nu b}}- {\tilde{\xi}_{\nu a}} {e_{\mu b}})-({\tilde{\xi}_{\mu b}} {e_{\nu a}}- {\tilde{\xi}_{\nu b}} {e_{\mu a}})]\nonumber
\end{eqnarray}
and
\begin{eqnarray}
{{B}_{\mu\nu ab}} &\rightarrow&   {{B}_{\mu\nu ab}}+D_\mu {{\xi}_{\nu ab}}-D_\nu {{\xi}_{\mu ab}}\nonumber 
\end{eqnarray}
where the covariant derivative $D_\mu$ is defined as: 
\begin{eqnarray}
D_\mu {{\xi}_{\nu ab}}\equiv\partial_\mu {{\xi}_{\nu ab}}+{\omega_{\mu a}}^{c}{\xi_{\nu cb}}+{\omega_{\mu b}}^{c}{\xi_{\nu a c}}
\end{eqnarray} 

Therefore the above action can be regarded as the $3+1$D generalization of $2+1$D topological gravity. Apparently, the beta function vanishes for $S_{\rm{Top}}$ and it is renormalizable. The argument is exactly the same as the $2+1$D case, where the counterterms, if any, are integrals of local gauge invariant functional and can not renormalize the above action. Another straightforward argument is that for compact gauge groups, all the terms in the above actions are actually quantized and the beta function must vanish\cite{aada1}. Similar to the 2+1D case, $e, \omega$ are dimension one operators while $B$ is dimension two operator. Finally, the equation of motion implies the vanishing of curvature and torsion tensors.
\begin{eqnarray}
&& R^{ab}=0,\quad T^a=0, \quad  \nabla \tilde{B}_a=-\epsilon_{abcd} e^b \wedge R^{cd}=0,\nonumber\\ && \nabla B_{ab}+\frac{1}{2}(\tilde{B}_a \wedge e_b-\tilde{B}_b \wedge e_a)=-\epsilon_{abcd} T^c\wedge e^d=0\nonumber
\end{eqnarray}

\textit{Quantization of topological gravity} --
Before discussing the possible connection with $3+1$D Einstein gravity, let us proceed the standard canonical quantization for the above topological gravity and explain its underlying physics. The Lagrangian density reads:
\begin{eqnarray}
&&\mathcal{L}_{\rm{Top}}={\Pi^i}_{ab}\partial_0{\omega_{i}}^{ab}+ {\pi^i}_a\partial_0{e_{i}}^{a}+\frac{1}{2} \epsilon^{ijk}B_{0i ab}{R_{jk}}^{ab}\nonumber\\&&+\frac{1}{2}\epsilon^{ijk}\tilde{B}_{0i a}{T_{jk}}^{a}+{e_0}^a(\nabla_i{\pi^i}_{a}+\frac{1}{2}\epsilon^{ijk}\epsilon_{abcd}{e_i}^b {R_{jk}}^{cd})\nonumber\\&&+{\omega_0}^{ab}\left[\nabla_i{\Pi^i}_{ab}+\frac{1}{2}({\pi^i}_a {e_i}^b-{\pi^i}_b {e_i}^a)\right]
\end{eqnarray}
where the canonical momentums of ${\omega_i}^{ab}$ and ${e_i}^a$ are defined as 
${\Pi^i}_{ab}=\frac{1}{2}\epsilon^{ijk}\epsilon_{a
bcd}{e_j}^c {e_k}^d +\frac{1}{2}\epsilon^{ijk}B_{jk ab}$ and ${\pi^i}_a=\frac{1}{2}\epsilon^{ijk}\tilde{B}_{jk a}$. 
 Canonical quantization requires:
\begin{eqnarray}
&&[{\omega_j}^{cd}(\vy),{\Pi^i}_{ab}(\vx)]=i\delta_j^i\delta_{ab}^{cd} \delta(\vx-\vy),\nonumber\\
&&[{e_j}^{b}(\vy),{\pi^i}_{a}(\vx)]=i\delta_j^i\delta_a^b \delta(\vx-\vy)\nonumber\\
&& \text{all other commutators}=0
\end{eqnarray}
with the following flat-connection   constraints:
\begin{eqnarray}
&&\frac{1}{2} \epsilon^{ijk}{R_{jk}}^{ab}=0,\quad\frac{1}{2}\epsilon^{ijk}{T_{jk}}^{a}=0 \nonumber\\ &&\nabla_i{\pi^i}_{a}=-\frac{1}{2}\epsilon^{ijk}\epsilon_{abcd}{e_i}^b {R_{jk}}^{cd}=0\nonumber\\&& \nabla_i{\Pi^i}_{ab}+\frac{1}{2}({\pi^i}_a {e_i}^b-{\pi^i}_b {e_i}^a)=0
\end{eqnarray}
Similar to the $2+1$D topological gravity, the phase-space to be quantized is exactly the solutions of above constraints divided by the group of gauge transformations generated by the constraints. The quantum Hilbert space is the flat connections of Poincare group modulo gauge transformations.(If we regard $e$ and $\omega$ as coordinates while $\pi$ and $\Pi$ as momentums.)Of course, in order to define an ultraviolet(UV) complete theory, it is much better to use the algebraic framework of tensor 2-category theory\cite{TCAT1,TCAT2}(It is well known that the $2+1$D Chern-Simons theory can be described by the algebraic tensor category theory.) 

In fact, the above constraints are exactly the same as the usual $BF$ theory of Poincare group, and the subtle difference only arises from the definition of physical observable corresponding to loop like excitation, namely, the Wilson surface operator. Let us rewrite the commutation relations in terms of $B,\tilde{B},e,\omega$:
\begin{eqnarray}
&&[{\omega_i}^{ab}(\vx),{B}_{jk cd}(\vy)]=i\epsilon_{ijk}\delta_{cd}^{ab} \delta(\vx-\vy),\nonumber\\
&&[{e_i}^{a}(\vx),\tilde{B}_{jkb}(\vy)]=i\epsilon_{ijk}\delta_b^a \delta(\vx-\vy),\nonumber\\
&&[{B}_{ij ab}(\vx),\tilde{B}_{klc}(\vy)]=i\epsilon_{abcd}({e_i}^d\epsilon_{jkl}-{e_j}^d\epsilon_{ikl})\delta(\vx-\vy), \nonumber\\
&&\text{all other commutators}=0. \label{com}
\end{eqnarray}

%Thus, we can construct the gauge invariant Wilson line operators as usual:
%\begin{eqnarray}
%{W_{e^a}}_{;\gamma}=\exp\left(i\oint_\gamma e^a\right); \quad {W_{\omega^{ab}}}_{;\gamma}=\exp\left(i\oint_\gamma \omega^{ab}\right)
%\end{eqnarray}
%However, the Wilson surface need to be modified as:
%\begin{eqnarray}
%&&{U_{\tilde{B}_a}}_{;\Omega}=\exp\left(i\int_\Omega \tilde{B}_a+i\int_V \epsilon_{abcd} e^b \wedge R^{cd} \right); \nonumber\\  &&{U_{B_{ab}}}_{;\Omega}=\exp\left[i\int_\Omega B_{ab}+\frac{i}{2}\int_V(\tilde{B}_a\wedge e_b-\tilde{B}_b\wedge e_a)\right]\nonumber
%\end{eqnarray}
%where $\Omega=\partial V$. It is straight forward to verify the gauge invariance of the above Wilson surface operators. 
In recent works, it has been pointed out that such modified commutation relations actually imply the nontrivial three-loop-braiding\cite{aada1,aada2,aada3} statistics among flux lines of gauge fields, which makes it different from the usual $BF$ theory of Poincare group with trivial three-loop-braiding statistics. 

\textit{Loop condensation and the emergence of Einstein gravity} -- 
To this point, one may wonder why we are interested in the $3+1$D topological gravity theory which is somewhat trivial. Here we conjecture that quantum gravity is actually controlled by a topological gravity fixed point and the classical space-time vanishes at extremely high energy scale. Therefore it is quite natural to expect the vanishing of curvature and torsion at that scale. Mathematically, $3+1$D TQFT can be described and classified by tensor 2-category theory and 
a possible way to generate interesting dynamics is condensing loops(flux lines in the context of topological gauge theory). 
%from the underlying tensor 2-cateogry theory. 
If we further assume that the condensed loop carries a nontrivial linking Berry phase\cite{BB1,BB2,BB3}, a $\int Tr(B\wedge B)$ type term can be induced. Let us consider the following term:
\begin{eqnarray}
S_\theta=-\frac{\theta}{2} \int B_{ab}\wedge B^{ab}
\end{eqnarray}
This term breaks the 2-form gauge symmetry as well as the translational gauge symmetry explicitly. A microscopic derivation of the above term from loop condensation is beyond the scope of this paper. Here we just introduce such a term phenomenologically to describe low energy dynamics and ignore all the microscopic details of loop dynamics, which is the analog of using massive gauge boson to describe Abelian Higgs phase and considering the infinite massive limit for Higgs boson. (More precisely, one can assume that the total action $S$ is consisting of two terms $S_{\rm{Top}}$ and $S_{\rm{Loop}}$ at UV scale. $S_{\rm{Loop}}$ describes the dynamics of closed loop and it can be approximated by $S_{\theta}$ in the loop condensed phase after taking the infinite massive limit for loop.) Remarkably, for small $\theta$, the total action $S=S_{\rm{Top}}+S_\theta$ is still power-counting renormalizable since $S_\theta$ only contains dimension four operators. %Actually the topological nature of $S_\theta$ suggests the vanishing of beta function for $S_{\theta}$ as well. 
A detailed calculation of beta functions will be presented elsewhere.  

%Now we will ignore all the high energy details for a quantum theory of gravity and just use $S$ phenomenologically as the low energy effective description. 
%Below we will show that the $\theta \rightarrow 0$ limit actually corresponds to the classical limit of $e$ and $\omega$ fields, and 
The classical equation of motion for the total action $S$ reads:
\begin{eqnarray}
&& B^{ab}=\frac{1}{\theta}R^{ab},\quad T^a=0, \quad 
\epsilon_{abcd} e^b \wedge R^{cd}= -\nabla \tilde{B}_a, \nonumber\\ &&\epsilon_{abcd} T^c\wedge e^d+\frac{1}{2}(\tilde{B}_a \wedge e_b-\tilde{B}_b \wedge e_a)=-\nabla B_{ab}
\end{eqnarray}
Insert the first two equations into the last equation, we have:
\begin{eqnarray} 
%&&\epsilon_{abcd} e^b \wedge R^{cd}=- \nabla \tilde{B}_a, \nonumber\\ &&
\frac{1}{2}(\tilde{B}_a \wedge e_b-\tilde{B}_b \wedge e_a)=-\frac{1}{\theta}\nabla R_{ab}=0
\end{eqnarray}
The above equation can be rewritten in a compact form as $\epsilon^{abcd} \tilde{B}_a\wedge e_b=0$, which further implies $\tilde{B}^a=0$. Thus, we eventually derive the vacuum Einstein-Cartan equation: 
\begin{eqnarray}
\epsilon_{abcd} e^b \wedge R^{cd}=0.
\end{eqnarray}

\textit{Einstein gravity as a non-commutative geometry} -- 
Now let us proceed the canonical quantization for the total action $S$. The total Lagrangian density reads:
\begin{widetext}
\begin{eqnarray}
\mathcal{L}&=&{\Pi^i}_{ab}\partial_0{\omega_{i}}^{ab}+ {\pi^i}_a\partial_0{e_{i}}^{a}+\frac{1}{2} \epsilon^{ijk}B_{0i ab}({R_{jk}}^{ab}-\theta{B_{jk}}^{ab})\nonumber\\&+&\frac{1}{2}\epsilon^{ijk}\tilde{B}_{0i a} {T_{jk}}^{a}+{e_0}^a(\nabla_i{\pi^i}_{a}+\frac{1}{2}\epsilon^{ijk}\epsilon_{abcd}{e_i}^b {R_{jk}}^{cd})+{\omega_0}^{ab}\left[\nabla_i{\Pi^i}_{ab}+\frac{1}{2}({\pi^i}_a {e_i}^b-{\pi^i}_b {e_i}^a)\right]
\end{eqnarray}
\end{widetext}
where the canonical momentum $\Pi_{ab}^i,\pi_a^i$ have the same definition as in $3+1$D topological gravity. By integrating out $B_{0iab}$ and $\tilde{B}_{0ia}$, we derive the following constraints:
\begin{eqnarray}
{B_{ij}}^{ab}=\frac{1}{\theta}{R_{ij}}^{ab} ,\quad  {T_{ij}}^{a}=0
\end{eqnarray}
We note that the torsion free condition arises as a quantum constraint instead of equation of motion here. This feature is very different from the usual Einstein-Cartan theory. 
The canonical quantization conditions Eq.(\ref{com}) imply the following noncommutative geometry:
\begin{eqnarray}
&&[{\omega_i}^{ab}(\vx),{R}_{jk cd}(\vy)]=i\theta\epsilon_{ijk}\delta_{cd}^{ab} \delta(\vx-\vy),\\
&&[{e_i}^{a}(\vx),\tilde{B}_{jkb}(\vy)]=i\epsilon_{ijk}\delta_b^a \delta(\vx-\vy),\nonumber\\ &&[{R}_{ij ab}(\vx),\tilde{B}_{klc}(\vy)]=i\theta\epsilon_{abcd}({e_i}^d\epsilon_{jkl}-{e_j}^d\epsilon_{ikl})\delta(\vx-\vy),\nonumber
\end{eqnarray}
In the semi-classical limit with, both $R$ and $e$ fields have weak quantum fluctuations while $\omega$ and $\tilde{B}$ fields have strong quantum fluctuations. A very interesting observation is that the small parameter $\theta$ enter the commutation relation and it will be very interesting to understand the relationship between $\theta$ and planck constant $\hbar$ in our future work. 
%If this is indeed the case, the loop condensation picture not only explains the original of classical space time, but also explain the origin of $\hbar$!  
%We see that the $\theta \rightarrow 0$ limit corresponds to the classical limit with commutative geometry. 
%For small but finite $\theta$, it is in principle possible to measure the non-commutative geometry between curvature and torsion tensors experimentally and this is a smoking gun to verify or falsify our theory.

To this end, we see that the nature of quantum gravity is the emergence of non-commutative geometry via loop condensation from an underlying topological gravity theory. We stress that $S_{\rm{Top}}$ with vanishing $R$ and $T$ describes the \textit{absolute vacuum} of our universe in the absence of classical space time, and it might provide a new route towards resolving the black hole singularity as well as the big bang singularity.
%Another interesting point is that since the planck constant $\hbar$ does not enter the (super) tensor 2-category theory(which is believed to be a UV-complete description for $3+1$D TQFT), the parameter $\theta$ might be related to the origin of $\hbar$ since the classical limit is usually achieved by taking $\hbar \rightarrow 0$. %In fact, the nontrivial gauge transformation Eq.(\ref{gauge}) also implies that $S_\theta$ breaks translational gauge symmetry and it might lead to the emergence of absolute time. 
%Unfortunately, we are not able to explore the precise relation between $\theta$ and $\hbar$ without a microscopic model of $S_{\theta}$ and we will leave this interesting problem to our future work. 

\textit{Cosmological constant term} --
Our construction for topological gravity action in $3+1$D can be easily generalized into the case with cosmological constant term:
\begin{eqnarray}
S_{\rm{Top}}^\prime&=&S_{\rm{Top}}+\frac{\Lambda}{4!}\int \epsilon_{a
bcd} e^a \wedge e^b \wedge e^c \wedge e^d  \nonumber\\
&=&S_{\rm{Top}}+\frac{\Lambda}{4!}\int d^4x \epsilon^{\mu\nu\rho\sigma} \epsilon_{a
bcd} e^a_{\mu} e^b_{\nu} e^c_{\rho} e^d_{\sigma}  
\end{eqnarray}
We only need to properly redefine the gauge transformation of translational symmetry:  
\begin{eqnarray}
{e_\mu}^a &\rightarrow & {e_\mu}^a+D_\mu f^a \equiv {e_\mu}^a+\partial_\mu f^a+{\omega_\mu}^{ab} f_b \nonumber\\
\tilde{B}_{\mu\nu a} &\rightarrow & \tilde{B}_{\mu\nu a}-\epsilon_{a
bcd} f^b {R_{\mu\nu}}^{cd}-\frac{\Lambda}{2}\epsilon_{a
bcd} (e_{\mu}^b e_{\nu}^c-e_{\nu}^b e_{\mu}^c) f^d \nonumber \\
B_{\mu\nu ab} &\rightarrow & B_{\mu\nu ab}-\frac{1}{2}(\tilde{B}_{\mu\nu a}f_b-\tilde{B}_{\mu\nu b}f_a)
\end{eqnarray}
Similar to the case without cosmological constant term, loop condensation will lead to Einstein-Cartan action with cosmological constant term, and the whole theory remains to be power-counting renormalizable.  

\textit{Super symmetric generalization} -- Finally, let us discuss the SUSY generalization of $3+1$D topological gravity. Similar to the $2+1$D topological gravity theory, we just need to introduce the gauge connection of super Poincare group and write the action as $\int sTr[ A\wedge A\wedge (dA+A\wedge A)]+\int sTr(B \wedge F)$. For example, for the $N=1$ case, we can just express $A$, $B$ and $F$ as:
\begin{eqnarray}
&&A_{\mu}\equiv \frac{1}{2}\omega_{\mu}^{ab} M_{ab}+e_\mu^a P_a+\bar{\psi}_{\mu\alpha} Q^\alpha \nonumber\\
&&B_{\mu\nu}\equiv \frac{1}{2}{B_{\mu\nu}}^{ab} M_{ab}+{\tilde{B}_{\mu\nu}}^a P_a+\mathfrak{B}_{\mu\nu\alpha} Q^\alpha \nonumber\\
&& F_{\mu\nu}\equiv
\frac{1}{2}{R_{\mu\nu}}^{ab} M_{ab}+T_{\mu\nu}^a P_a+\bar{R}_{\mu\nu\alpha} Q^\alpha
\end{eqnarray}
Here $\bar{R}_{\mu\nu\alpha}$ is the super curvature tensor defined as
$\bar{R}_{\mu\nu\alpha}=D_{\mu} \bar{\psi}_{\nu\alpha}-D_{\nu} \bar{\psi}_{\mu\alpha}$ where $D_{\mu}$ is the covariant derivative for spinon fields.
However, as fermionic loops(flux lines) can not be condensed, super symmetry breaking already happens at very high energy scale when bosonic loops condense and classical space-time emerges. Thus, the super curvature $\bar{R}_{\mu\nu\alpha}$ always vanishes and the semiclassical limit of $3+1$D quantum gravity can still be described by $S$ at low energy. However, the SUSY generalization might provide us a natural way to extend the our model to include fermionic matter fields.%Since Chern-Simons theory of super Lie algebra is well understood in $2+1$D, we believe that a super tensor category theory can be properly defined in principle and can be generalized into higher dimensions as well.

\textit{Topological gravity in arbitrary dimensions and the emergence of $3+1$D space-time} -- Before conclusion, let us generalize topological gravity theory into arbitrary dimensions with the following gauge invariant action: 
\begin{eqnarray}
S_{\rm{Top}}&=&\frac{1}{n-2}\int \epsilon_{a
b a_3\dots a_n} R^{ab}\wedge e^{a_3} \wedge \dots\wedge e^{a_n}\nonumber\\
&+&\int C_{ab} \wedge R^{ab}+\int \tilde{C}_{a} \wedge T^{a}
\end{eqnarray}
where $C$ and $\tilde{C}$ are $n-2$ forms. Interestingly, we see that it is only possible to introduce $\int C \wedge C $ type term for four dimensional space-time. Thus, we may start with a model describing topological gravity in all dimensions(e.g. topological nonlinear sigma model of the Poincare group classifying space) and condense the loop, only the four dimensional vielbein field admits a semi-classical limit that defines the classical space-time!
%Similarly, the $n$-dimensional Einstein-Cartan action is not a well defined quantum field theory and we need to define the corresponding topological gravity theory to make it perturbative renormalizable.(Of course, for UV-completion, we still need to invent a super fusion n-category theory description.)

\textit{Conclusions and discussions} -- In conclusion, we propose a topological paradigm to understand $3+1$D quantum gravity. In particular, we generalize Edward Witten's $2+1$D topological gravity theory into arbitrary dimensions. In $3+1$D, by condensing loops(flux lines), we find a semi-classical limit where the Einstein-Cartan equation emerges.(In the absence of matter fields.) Our approach can be generalized into the case with cosmological constant term. In fact, it is well known that starting from a topological BF theory of gauge group $G$ with action $S_{\rm{top}}=\int Tr (B\wedge F)$, condensing the loops by introducing a mass term $S_g=g\int d^4x Tr (B^2)$(we consider flat space-time background here for simplicity) is another way to derive a gauge theory with a Maxwell term $\frac{1}{g} \int d^4x Tr (F^2)$(by integrating out the $B$ fields and regarding $g$ as the coupling constant). Thus, we argue that the concept of condensing loops from a topological gauge theory(which can be rigorously defined as topological nonlinear sigma model in classifying space of the corresponding gauge group\cite{DWmodel}) might provide us a unified description for both gauge theory and gravity. 
%Apparently, how to use tensor 2-category theory to precisely describe the complete details of the loop condensation will be a promising future direction. 
%Furthermore, loop like extensive object could be another candidate for dark matter.

\textit{Acknowledgments} --
We thank S T Yau for invitation to visit Center of Mathematical Sciences and Applications at 
Harvard University where this work was initialized. We also thank Dvide Gaiotto and Kevin Costello for critical reading and useful comments.
We acknowledge start-up support from Department of Physics, The Chinese University of Hong Kong, Direct Grant No. 4053224 from The Chinese University of Hong Kong and the funding from RGC(No.2191110).

\end{document}